# Extreme Microlensing

**Andrew Gould**[1]

Dept of Astronomy, Ohio State University, Columbus, OH 43210

e-mail gould@payne.mps.ohio-state.edu

## Abstract

Extreme microlensing events, defined as events with maximum magnification $A_{\max} \gtrsim 200$ are a potentially powerful probe of the mass spectrum and spatial distribution of objects along lines of sight toward the Galactic bulge. About 75 yr$^{-1}$ such events are expected for main-sequence sources with $I_0 < 19$. For many of these it is possible to measure both a "proper motion" and a "parallax" which together would yield individual mass, distance, and transverse-speed determinations of the lensing object. The proper motion is determined from finite-source effects when the lens transits, or nearly transits the source. The parallax is determined by observing the difference in the light curve as seen from two Earth observatories separated by about 1 Earth radius, $R_\oplus$. The size of the parallax effect is $\sim A_{\max} R_\oplus / \tilde{r}_e$ where $\tilde{r}_e$ is the projected Einstein radius. This can be of order 1%. Detection of candidate events requires a pixel-lensing search of the entire bulge once per day, preferably by at least two observatories on different continents. Follow-up observation must be carried using optical/infrared photometry, with short (e.g. 1 minute) exposures on small ($\gtrsim 1$ m) telescopes. Extreme microlensing observations toward the Large Magellanic Cloud do not appear feasible at the present time.

Subject Headings: Galaxy: structure – gravitational lensing – stars: mass function

---

## 1. Introduction

Three groups are presently searching for microlensing events toward the Galactic bulge, OGLE (Udalski et al. 1994), MACHO (Alcock et al. 1995a, 1996a) and Duo (Allard 1996). A fourth group, EROS (Aubourg et al. 1993, Ansari et al. 1996) will initiate such a search shortly. Events detected in this direction probe the mass content of Galactic disk (Paczyński 1991; Griest et al. 1991) as well as the bulge itself (Kiraga & Paczyński 1994). For microlensing by a point source, the observed flux $F(t)$ from a lensed source star is given by $F(t) = F_0 A(t)$ where $F_0$ is the flux of the unlensed source and (Paczyński 1986)

$$A[x(t)] = \frac{x^2 + 2}{x(x^2 + 4)^{1/2}}, \qquad x(t) = [\omega^2(t - t_0)^2 + \beta^2]^{1/2}. \tag{1.1}$$

Here $\omega^{-1}$ is the time scale of the event, $t_0$ is the time of maximum, and $\beta$ is the impact parameter normalized to the angular Einstein radius, $\theta_e$,

$$\theta_e = \left(\frac{4GM}{c^2 D}\right)^{1/2}, \qquad D \equiv \frac{D_{\rm ol} D_{\rm os}}{D_{\rm ls}}, \tag{1.2}$$

where $M$ is the mass of the lens and $D_{\rm ol}$, $D_{\rm ls}$, and $D_{\rm os}$ are the distances between the observer, lens, and source. Of the three lensing parameters which can be extracted from a lensing event [cf. eq. (1.1)], only the time scale is related to the physical parameters of the lens,

$$\omega = \frac{v}{D_{\rm ol} \theta_e}, \tag{1.3}$$

where $v$ is the transverse speed of the lens relative to the observer-source line of sight. The other two parameters, $t_0$ and $\beta$, simply reflect the geometry of the event.

One would like to use the observed lensing events to learn about the details of the lens population. For example, is this population fully accounted for by the known populations of luminous stars? What is the mass spectrum of the lenses? What is their distribution along the line of sight? What are their kinematic properties? Because the one observable $\omega$ is a complicated combination of the physical properties of the lens, it is difficult to obtain unambiguous answers to these questions. Zhao, Spergel, & Rich (1995) and Han & Gould (1996a) estimated the mass spectrum from the observed distribution of time scales by assuming that the sources and lenses have velocity and spatial distributions like those of observed stars. Han & Gould (1996a) found that the inferred mass spectrum is inconsistent at the $5\sigma$ level with that of nearby stars as determined by Gould, Bahcall, &

Flynn (1996) using *Hubble Space Telescope (HST)* observations. If confirmed by continuing observations, this would be an extremely intriguing result. Nevertheless, the adopted approach is fundamentally limited both by its statistical nature and by its dependence on unverifiable assumptions about the phase space distribution of the lenses. One would like to be able to measure $M$, $D_{\rm ol}$, and $v$ for each individual lens, or at least for a representative subsample of events.

It would be possible to determine individual masses provided one could somehow measure $\theta_e$ [cf. eq. (1.2)] and $\tilde{r}_e$, the Einstein radius projected onto the observer plane,

$$\tilde{r}_e = D\theta_e = \left(\frac{4GMD}{c^2}\right)^{1/2}. \tag{1.4}$$

From equations (1.2) and (1.4), one finds

$$M = \frac{c^2}{4G}\tilde{r}_e\theta_e. \tag{1.5}$$

In fact, since $D_{\rm os}$ is typically known to within $\sim 10\%$ simply from the source's membership in the bulge, one also gets a good estimate of the position and transverse speed of the lens,

$$D_{\rm OL} = \left(\frac{1}{D_{\rm OS}} + \frac{\theta_e}{\tilde{r}_e}\right)^{-1} \qquad v = \frac{\omega}{\tilde{r}_e^{-1} + (\theta_e D_{\rm os})^{-1}}. \tag{1.6}$$

Measurement of $\tilde{r}_e$ is often called a "parallax" because it is found by observing the lensing event from two different positions in the observer plane. Measurement of $\theta_e$ is often called a "proper motion" because the product $\mu = \omega\theta_e$ is the angular speed of the lens relative to the source.

There is no lack of ideas for measuring parallaxes and proper motions for special rare classes of events. For example, for long events the position of the Earth changes enough during the event to allow a parallax measurement (Gould 1992; Alcock et al. 1995b). However, while the long events are an interesting sub-class, they are by definition unrepresentative of the lenses as a whole. Moreover, parallax measurements do not by themselves permit determination of the mass without a simultaneous proper motion measurement, and the fraction of long events for which such measurements are possible is small. To be useful as probes of the lens mass spectrum, what is required is that *both* quantities be measured for a *representative sample* of events.

One approach is to obtain parallaxes using a satellite in solar orbit (Refsdal 1966; Gould 1994b,1995b; Gaudi & Gould 1996), and proper motions from finite source effects for small $\theta_e$ and from optical interferometry for large $\theta_e$. With next generation instruments, this approach could yield $\sim 35$ mass measurements per year with no serious selection bias (Gould 1996b).

Here I discuss another approach which, while substantially less effective than the one just described, could be initiated much earlier.

## 2. Extreme Microlensing Events

The basic idea is to measure both $\tilde{r}_e$ and $\theta_e$ for a very special, yet nearly representative sub-class of events: the extreme magnification events (EMEs). EMEs are events with maximum magnifications

$$A_{\max} \gtrsim Q, \qquad (2.1)$$

where $Q$ is a large number, typically $Q \sim 200$. For equation (2.1) to hold, two physical conditions must be satisfied:

$$\beta \lesssim Q^{-1}, \qquad \theta_* \lesssim Q^{-1}\theta_e, \qquad (2.2)$$

where $\theta_*$ is the angular radius of the source star. The first condition restricts the geometry of the event, while the second restricts the class of source stars. The value of $Q$ (i.e., the selection function) has a well-understood dependence on the physical characteristics of the lens, which accounts for the above description of EMEs as "nearly representative" (see § 3).

### 2.1. EME Parallaxes

Because of parallax, microlensing events appear slightly different when viewed from different observatories on Earth (Holz & Wald 1996). Just as with satellite parallaxes (Gould 1994b), the events will have different impact parameters $\beta$ and $\beta'$ and different times of maximum $t_0$ and $t_0'$. The difference can be combined into a single vector $\Delta\mathbf{x}$,

$$\Delta\mathbf{x} = (\omega\Delta t, \Delta\beta), \qquad (2.3)$$

where $\Delta t \equiv t_0' - t_0$ and $\Delta\beta \equiv \beta' - \beta$. Let the separation between the observatories (projected onto the plane perpendicular to the line of sight) be $d_{\text{sep}}$. Then, if $\Delta\mathbf{x}$ can be measured, one can determine $\tilde{r}_e$,

$$\tilde{r}_e = \frac{d_{\text{sep}}}{\Delta x}, \qquad \Delta x \equiv |\Delta\mathbf{x}|. \qquad (2.4)$$

Of course, since typically $\tilde{r}_e \sim \mathcal{O}(\text{AU})$ and $d_{\text{sep}} \sim R_\oplus$ where $R_\oplus$ is the radius of the Earth, $\Delta x$ is incredibly small: $\Delta x \sim R_\oplus/\text{AU} \sim 1/25{,}000$. Not surprisingly, the microlensing community greeted this suggestion with some skepticism, and Holz & Wald (1996) themselves made no claims that the effect could actually be observed, only that photon statistics alone do not preclude such observations.

For EMEs, however, such Earth-based parallaxes are within the range of present capabilities. This is because the observable effects do not scale as $\Delta \mathbf{x}$, but as $\Delta \mathbf{x}/x$. Since $x \sim Q^{-1}$ near the peak, EME parallax effects are $\mathcal{O}(1\%)$. To make a quantitative analysis, I assume that photometry is limited by systematic effects to some fixed fractional accuracy $\sigma$ (rather than being photon limited), but that these errors are uncorrelated. The event is observed from $t_0 - T$ to $t_0 + T$ at a rate $N\omega\beta^{-1}$. That is, the observations are carried out $N$ times per "effective time scale", $t_{\rm eff}$, where

$$t_{\rm eff} \equiv \frac{\beta}{\omega}. \tag{2.5}$$

I then find that the errors $\delta t_0$ and $\delta \beta$ in the measurements of $t_0$ and $\beta$ are given by (e.g. Gould 1995a),

$$\frac{\delta t_0}{t_{\rm eff}} = \frac{\sigma}{\{N[\Theta - \sin(2\Theta)/2]\}^{1/2}}, \quad \frac{\delta \beta}{\beta} = \frac{\sigma}{\{N[\Theta + \sin(2\Theta)/2]\}^{1/2}}, \quad \tan\Theta \equiv \frac{T}{t_{\rm eff}}. \tag{2.6}$$

For simplicity, I henceforth assume that the observations can be carried out long enough so that both errors $\lesssim N^{-1/2}\sigma$. For typical events seen toward the bulge, $\omega^{-1} \sim 10\,{\rm days}$ (Alcock et al. 1995a). Hence the effective time scale for an EME with $\beta^{-1} \sim 200$ is $t_{\rm eff} \sim 1\,{\rm hr}$. Assuming one could make one observation per minute each with fractional accuracy $\sigma = 1\%$, then $\delta t_0/t_{\rm eff} \sim \delta\beta/\beta \sim 0.13\%$, implying an accuracy in the determination of $\Delta x$ of $\sim \beta/550 \sim 10^{-5}$. Recall that the typical scale of interest is $\Delta x \sim R_\oplus/{\rm AU} \sim 4 \times 10^{-5}$. For lower $\beta$, but the same accuracy and rate of observations, the determination improves as $\beta^{1/2}$.

These results show that Earth-based parallax measurements of EMEs are generally within the range of present technology.

## 2.2. EME Proper Motions

When the lens transits the source, the light curve deviates from the point-source form (1.1). One can then measure $x_*$,

$$x_* \equiv \frac{\theta_*}{\theta_e}, \tag{2.7}$$

the value of $x$ when transit occurs (Gould 1994a; Nemiroff & Wickramasinghe 1994; Witt & Mao 1994). If $\theta_*$ is known (as it usually is from the dereddened color and magnitude and Stefan's law), then one can determine $\theta_e = \theta_*/x_*$. If the lens comes close to the source but does not transit, there is still a fractional deviation from the point-source formula $\Delta A/A \sim (\Lambda/8)(x_*/x)^2$, where $\Lambda$ is the second radial moment of the source normalized so that $\Lambda = 1$ for a uniform disk. Unfortunately, with single-band photometry one cannot put this effect to use because it cannot be distinguished from a slight shift in $\beta$ (Gould & Welch 1996). However, since stars are limb-darkened by different amounts in different bands, near transits give rise to color effects which can be measured (Witt 1995). Specifically, Gould & Welch (1996) find $\Lambda^H - \Lambda^V = 0.07$, allowing measurement of $\theta_e$ for $\beta \lesssim 2x_*$. Since EMEs typically fall in or near this range, it will often be possible to measure their proper motions.

### 2.3. Combined Parallaxes and Proper Motions

At first sight, it may appear that the very condition required to measure $\theta_e$ (transit or near-transit of the source) would make measurement of $\tilde{r}_e$ impossible. In fact, the majority of mass measurements are not severely affected by this potential problem. Consider first an event with $\beta = 1/200$ and $x_* = 1/300$. At the peak of the event, the perturbation due to finite size is $\Delta A/A = (\Lambda/8)(x_*/\beta)^2 \sim 5\%$ (where I have assumed $\Lambda = 0.9$). Since this is several times the change in $A$ due to parallax ($\sim \Delta\beta/\beta \sim 1\%$), one might worry that it would render the parallax shift unobservable. In fact, since the finite-size effect (at fixed source-lens separation) is identical for the two observers, the difference in their observed maximum magnifications still accurately measures $\Delta\beta$. The finite-source effect would lead to $\sim 5\%$ fractional error in the estimate of $\Delta\beta/\beta$ if left uncorrected, but even the correction to this minor systematic error is not difficult to determine once the size of the source is measured.

If the lens actually transits the source, $\beta < x_*$, then the situation is more complicated. In this case, one could restrict attention to those portions of the light curve where $x \gtrsim x_*$, for which the light curve is either unaffected by finite-source effects or the corrections due to these effects are well determined. (As in the previous example, one is interested only in the difference between the two curves, so the corrections play a minor role.) I assume in this case there are $N$ measurements per stellar crossing time $t_* \equiv x_*/\omega$, each with accuracy $\sigma$, and that the measurements are carried out with over a symmetric interval of half width $T$. I then find (see e.g. Gould 1995a)

$$\frac{\delta t_0}{t_*} = \frac{\sigma}{N^{1/2}} \left\{ \frac{x_*}{\beta} \left[ (\Theta_f - \Theta_i) - \left( \frac{\sin 2\Theta_f}{2} - \frac{\sin 2\Theta_i}{2} \right) \right] \right\}^{-1/2}, \qquad (2.8)$$

$$\frac{\delta\beta}{x_*} = \frac{\sigma}{N^{1/2}}\left\{\frac{x_*}{\beta}\left[(\Theta_f - \Theta_i) + \left(\frac{\sin 2\Theta_f}{2} - \frac{\sin 2\Theta_i}{2}\right)\right]\right\}^{-1/2}, \qquad (2.9)$$

where $\tan\Theta_f \equiv \omega T/\beta$ and $\cos\Theta_i \equiv \beta/x_*$. For $\beta \ll x_*$, these equations have the limiting forms

$$\frac{\delta t_0}{t_*} \to \frac{\sigma}{(N/2)^{1/2}}, \qquad \frac{\delta\beta}{x_*} \to \frac{\sigma}{(3N)^{1/2}}\frac{\beta}{x_*}, \qquad (\beta \ll x_*) \qquad (2.10)$$

Equation (2.10) shows that if $\Delta t$ can be measured in a marginal transit event ($\beta = x_*$) with a given accuracy, then approximately the same accuracy can be achieved for all transit events ($\beta < x_*$). However, the accuracy of the measurement of $\Delta\beta$ deteriorates linearly with impact parameter as the impact parameter falls well below the source size. In § 6, I discuss the possibility of compensating for this loss of information about $\Delta\beta$ by making observations from a third site.

## 2.4. Marginal Transit Events Are Optimal

From the foregoing discussion, it is clear that the best events are those for which the lens just transits the limb of the star, $\beta = x_*$. For larger $\beta$, the parallax effect declines inversely as $\beta$ and for $\beta > 2x_*$ the proper motion cannot be measured. On the other hand, for smaller $\beta$, the measurement of $\Delta\beta$ becomes more difficult. Even if one compensates for this problem by making observations from a third site (see § 6) parallax measurements are still no more precise than for marginal transits. Thus, marginal transit events allow us to understand the fundamental limits of the technique.

The maximum parallax effect occurs at transit and is given by $\Delta x/x_*$ which may be evaluated,

$$\frac{\Delta x}{x_*} = \frac{d_{\rm sep}/\tilde{r}_e}{\theta_*/\theta_e} = \frac{d_{\rm sep}}{R_*}\frac{D_{\rm ls}}{D_{\rm ol}}, \qquad (2.11)$$

where $R_* = D_{\rm os}\theta_*$ is the physical radius of the source, and where in the last step I have made used $\tilde{r}_e = D\theta_e$ from equations (1.2) and (1.4). As I show in § 4, the typical source stars for EMEs are solar-type stars, or slightly fainter. Assuming the observatories are about 1 Earth radius apart, equation (2.11) becomes

$$\frac{\Delta x}{x_*} = \frac{R_\oplus}{R_\odot}\frac{D_{\rm ls}}{D_{\rm ol}} \sim 0.01\,(z^{-1} - 1), \qquad (2.12)$$

where $z \equiv D_{\rm ol}/D_{\rm os}$ is the fractional distance of the lens to the source. Hence to measure the mass of a disk lens ($z \sim 0.5$) requires detection of a 1% effect and for a bulge lens ($z \gtrsim 0.75$) requires detection of $\lesssim 0.3\%$ effect. While the exact threshold of the experiment cannot be determined without a better understanding of the limits to the photometric accuracy, it is clear that bulge events with sufficiently small lens-source separation will be beyond the limit. I call this limit $z_{\max}$. I discuss the effect of this limit on the selection function in § 3, and possible methods for extending it in § 5.

## 3. Selection Function

Let $\mathcal{S}(M, D_{\mathrm{ol}}, z)$ be the fraction of lensing events with parameters $M$, $D_{\mathrm{ol}}$, and $z = D_{\mathrm{ol}}/D_{\mathrm{os}}$ that have measurable parallaxes and proper motions. As discussed at the end of the previous section, the measurement precision sets a limit $z_{\max}$ such that for $z < z_{\max}$, parallax cannot be measured. The next most important selection effect is that to measure proper motions, the impact parameter must satisfy $\beta < 2x_*$. That is, $\mathcal{S} \propto \theta_*/\theta_e$. Since parallax measurements generally require small source stars, I initially assume that $\theta_*$ is fixed. (I relax this assumption below.) Hence,

$$\mathcal{S}(M, D_{\mathrm{ol}}, z) \propto \theta_e^{-1} \Theta(z_{\max} - z) \propto \left(\frac{M}{D}\right)^{1/2} \Theta(z_{\max} - z), \qquad (3.1)$$

where $\Theta$ is a step function.

While equation (3.1) is important for understanding the relation between the events with measured masses and the full ensemble of events, it is not the most useful form of the selection function. What is fundamentally of interest is not the distribution of parameters for the ensemble of lenses, but the distribution for the underlying populations of objects that give rise to the events. The lensing events are themselves a biased sample of the underlying population. They occur with relative frequency $\mathcal{F}$ proportional to their (one-dimensional) cross section and transverse speed, i.e., $\mathcal{F} \propto \theta_e D_{\mathrm{ol}} v$. Hence the fraction of all objects whose mass can be measured is

$$\mathcal{F} \times \mathcal{S} \propto D_{\mathrm{ol}} \bar{v}(D_{\mathrm{ol}}) \Theta(z_{\max} - z), \qquad (3.2)$$

where $\bar{v}(D_{\mathrm{ol}})$ is the mean transverse speed of objects at distance $D_{\mathrm{ol}}$. For the simplest models (see e.g. fig. 8 from Han & Gould 1995), one expects $\bar{v} \propto D_{\mathrm{ol}}$, in which case $\mathcal{F} \times \mathcal{S} \propto D_{\mathrm{ol}}^2 \Theta(z_{\max} - z)$. This result implies that EME mass measurements heavily favor more distant populations until the limit of parallax detection is reached close to the bulge. It therefore emphasizes the importance of pushing that limit as far as possible. See § 5.

While equation (3.2) reflects the most important selection effects, there are other effects which induce some additional minor modifications. First, higher mass lenses are slightly favored relative to equation (3.2) (which has no mass dependence). To see this, consider two masses with $M_1 = 4\,M_2$, both at the same distance $D_{\rm ol}$. For illustration, assume that the parallax and proper motion of $M_2$ are just measurable when $\beta_2 = 2x_* = 2\theta_*/\theta_{e,2}$ for a fiducial source star, $I_0 = 19$. In the above analysis, it was assumed that for the same star, the larger mass would have measurable proper motion only if $\beta_1 < \theta_*/\theta_{e,1}$, which is half as great. This is true, provided the source is the same. However, if $M_1$ were lensing a source star with twice the radius of the fiducial source, proper motions would be measurable to twice the impact parameter. Such larger stars are accessible to $M_1$ (but not $M_2$) because $\theta_{e,1}$ is larger and so the parallax effect is larger at fixed angular separation.

The reason that this is not a major effect is the steepness of the luminosity function which scales inversely with luminosity (when binned in magnitude intervals). Assuming that all stars have the same temperature (which is approximately true near the turn off), then stars with two times greater radius are four times less numerous. Hence, even for more massive lenses, most of the events with measurable proper motions will be near the magnitude limit.

A similar effect also favors nearby lenses because these also have larger Einstein rings: $\theta_e \propto D^{-1/2}$. The effect is likewise small.

## 4. Event Rate and Detection Strategy

A fraction $Q^{-1}$ of all events that have $A_{\max} > Q$, where I have for the moment ignored finite-size effects. Since the present detection rate is $\mathcal{O}(100)\,{\rm yr}^{-1}$, this would seem to imply that there would be $\lesssim 1$ event per year for $Q \sim 200$. However, the present detection strategy is not optimized for finding EMEs. Here, I show how an aggressive search could yield $\sim 75$ EMEs per bulge season.

Consider a main sequence star in the bulge with $I_0 \sim 19$. If this star were magnified by a factor $A \sim 200$, it would have a dereddened apparent magnitude $I_{0,A=200} = 13.5$, i.e., it would be brighter than most clump giants. Hence, at least near the peak, it would be as easy as a giant to photometer. One could hope to achieve 1% photometry or even better on such stars. Suppose that the star lay behind several magnitudes of extinction. The photometry problems induced by crowding would not change relative to the extinction-free case since all neighbors would suffer the same extinction. The photometry would be degraded only if there were insufficient photon statistics. Assuming $1''$ seeing and a sky brightness of $I = 19.6\,{\rm mag\,arcsec}^{-2}$, photon statistics predict errors of $\lesssim 1\%$ for a one minute exposure on a 1 m telescope at $I = 17$. As I have earlier discussed (Gould 1995c), there are $\sim 10^7$ giants ($I_0 < 15$) over an $82\,\mathrm{deg}^2$ area of the bulge with extinction $A_I < 3.5$. Using the bulge luminosity function measured by Light, Baum, & Holtzman (1996), I estimate that there are $\sim 2.5 \times 10^8$ stars with $I_0 < 19$ in the same region. Assuming an average optical depth $\tau \sim 3\times 10^{-6}$ (Alcock et al. 1996a), and a mean time scale $\langle \omega^{-1} \rangle \sim 20\,\mathrm{days}$, this leads to an estimate of $\sim 75\,\mathrm{yr}^{-1}$ events for $Q = 200$.

It is clearly impossible to identify these events using current search techniques which rely on following the light curves of stars recognized as such in a template image. Since the templates contain few if any of the $I_0 = 19$ stars, lensing events of such stars cannot be detected. Instead, one must make a pixel-lensing search of the type currently being carried out toward M31 (Crotts 1992; Baillon et al. 1993; Tomaney & Crotts 1996). In M31, there are many unresolved stars per pixel. One therefore subtracts a reference image from the current image to find *changes* in the brightness of individual stars. These changes appear as isolated point spread functions (PSFs) on an otherwise flat difference frame. In M31, pixel lensing is the only way to search for lensing events because there are essentially no resolved stars. On the other hand, it has not appeared necessary in the bulge or the Large Magellanic Cloud (LMC) because these fields contain many resolved stars. Note however that Melchior (1995) has made an initial attempt to find lensing events of unresolved stars in the LMC using pixel lensing, and efforts are continuing to develop this technique in fields with both resolved and unresolved stars. Pixel lensing is not as simple for the bulge as it is for M31 because the resolved stars in the field leave significant residuals in the difference images. To understand this problem concretely, consider a lensing event of an $I_0 = 19$ source with $\omega^{-1} = 10\,\mathrm{days}$ that is destined to become an EME. One day before maximum, it will have $I_{0,A=10} = 16.5$. While still about 5 times fainter than a giant, it would be substantially brighter than the net residuals from giants and of course would have a characteristic PSF shape which the residuals would not. Thus, it is likely that it could be recognized assuming that there was enough signal to noise. For the most heavily extincted regions under consideration, $A_I = 3.5$, the star would have $I = 20$, and so would be detectable with signal-to-noise ratio $\sim 25$ if it were on a blank field (assuming 5 minute observations on a 1 m telescope in $1''$ seeing). Whether it could actually be detected amidst the bulge-star residuals would depend on how well the subtraction worked. In any event, events in regions with $A_I < 2.5$ would very likely be detectable, and these include most of the available bulge field.

In brief, an aggressive pixel lensing search with a 1 m telescope and a 1 deg$^2$ camera, such as now is being commissioned by the EROS collaboration (M. Spiro, 1995 private communication), could cover the bulge each night with adequate depth to detect most events, weather permitting. There would be a substantial improvement in the detection rate if the bulge were covered from two continents. In this case one would benefit not only from reduced weather-induced gaps, but would also be more likely to expose when the object was bright enough to be detected but had not yet reached maximum. However, substantial improvements in the speed and efficiency of the real-time alert system would be required to enable the follow-up observations to begin before maximum.

## 5. Follow-Up Photometry

To obtain both parallaxes and proper motions, accurate photometry is required in two bands, preferably one optical, one infrared. The reason is that parallax measurements deteriorate rapidly for $\beta < x_*$, while proper motion measurements are impossible for $\beta > x_*$ unless there is photometry in two bands. It is possible to evade the parallax-measurement problem that arises at low impact parameters, but as I discuss in § 6, this evasion itself introduces significant logistical difficulties. Hence, the first requirement is to put specialized cameras equipped with dichoic beam-splitters (preferably optical/infrared) on telescopes dedicated to microlensing follow-up observations on several continents.

There are already two networks of observers currently engaged in follow-up photometry of ongoing microlensing events seen toward the bulge, PLANET (Albrow et al. 1996) and GMAN (Pratt et al. 1996). The primary objective of these networks is to find light-curve deviations that would be the signature of planets (Mao & Paczyński 1991; Gould & Loeb 1992). Like the EME observations proposed here, the planet searches require quick response to alerts and a high frequency of observations. And planet searches would benefit greatly from optical/infrared photometry (Gould & Welch 1996). In fact, such a camera has already been designed for this purpose and is the subject of an active proposal (D. Depoy 1996, private communication). Moreover, there is considerable interest in expanding the planet search. Since the planet search and the EME follow-up require similar instruments and modes of observation, it would be natural to combine the two.

A major goal of the follow-up photometry is to minimize the errors. Recall from § 2.4 that one typically expects the size of the parallax effect to be $\Delta x / x_* \sim 0.01 \, (z^{-1} - 1)$, and recall from § 2.1 that to measure this effect to $\sim 20\%$ accuracy requires the same order of precision in each 1 minute exposure, i.e., $1\% \, (z^{-1} - 1)$. Thus, if the measurement accuracy is limited to $\sigma \sim 1\%$, the mass measurements will reach only to $z_{\max} \sim 0.5$, that is, half way to the Galactic center. If the accuracy is $\sigma \sim 0.3\%$, then $z_{\max} \sim 0.75$ which would include most disk as well as some bulge events. If $\sigma \sim 0.15\%$, then events with $D_{\mathrm{ls}} \gtrsim 1\,\mathrm{kpc}$ will be accessible, which would give good sensitivity to bulge lenses.

The conventional wisdom is that 1% photometry is the limit for crowded fields, regardless of the signal-to-noise ratio. This view is borne of extensive experience with PSF-fitting of globular clusters and other crowded fields. Lensing searches have also used PSF-fitting as have all the follow-up searches. Measuring the mass of bulge lenses using EMEs will require another approach to photometry. Pixel-lensing techniques may provide the answer to this problem. I mentioned in § 4 that pixel lensing would be required to find the EMEs in the first place. However, the initial pixel-lensing search and the pixel-lensing follow-up observations have very different requirements and very different possibilities. In the initial search, a $10\,\sigma$ detection (and hence 10% photometry) would be quite adequate, while $< 1\%$ photometry is needed in the follow-up to improve on current techniques. On the other hand, the initial searches are driven to the largest pixel sizes consistent with Nyquist sampling in order to cover the largest angular area in the shortest time. Large pixels seriously degrade pixel-lensing photometry unless, as with the *Hubble Space Telescope (HST)* the pointing is extremely good (Gould 1996a). The follow-up observations are under no pressure toward large pixels and in fact several partners in PLANET and GMAN obtain highly oversampled data. These ongoing follow-up observations would make an excellent test bed for refining pixel-lensing techniques in fields containing resolved stars. If such refinements are successful, mass measurements for EMEs can be extended to lenses closer to the bulge. Otherwise they will be restricted to disk objects.

## 6. Degeneracies

EME parallaxes are in principle subject to the same two degeneracies that affect space-based parallaxes. First, the source positions as seen by the two observers can be on the same or opposite side of the lens, which leads to a two-fold degeneracy in the size of the Einstein ring (see figs. 1a and 1b from Gould 1994b). Second, there are two possible orientations of the source motion, which leads to a two-fold degeneracy in the inferred direction of the transverse velocity (see figs. 1a and 1d from Gould 1994b). However, the first degeneracy is almost always resolved for EMEs, and the second can be resolved in some cases but in any event is not important.

To see why the first degeneracy is not a major problem, consider an event generated by an object with $M = 0.3\,M_\odot$, $v = 150\,{\rm km\,s^{-1}}$, and $D_{\rm ol}/D_{\rm os} = 0.75$. And suppose that the parallax measurement yields $\Delta\beta/\beta \sim \omega\Delta t/\beta \sim 0.005$ based on the assumption that the source is on the same side of the Einstein ring. If the source were now assumed to be on the opposite side, then the inferred $\Delta\beta$ would be a factor $\sim 400$ larger, implying a larger $\Delta x$ and hence a smaller $\tilde{r}_e$ by a factor $\sim 280$. Using equation (1.6) one finds that the inferred transverse speed would then be $v \sim 2\,\mathrm{km\,s^{-1}}$ and the inferred distance $D_{\mathrm{ol}} \sim 20\,\mathrm{pc}$. For small distances and speeds, the cumulative event rate distribution $\propto v^3 D_{\mathrm{ol}}^2$, so the *a priori* probability of such an event is extremely low. For the transverse velocity to be so nearly perpendicular to the observatory separation vector that $\Delta\beta/(\omega\Delta t) = 400$ is even more improbable. Finally, the acceleration of the Earth ($\sim 0.5\,\mathrm{km\,s^{-1}\,day^{-1}}$) would produce easily observable effects over the course of a day unless the geometry were exceptionally unfavorable. As a practical matter, this form of degeneracy is therefore excluded.

The second form of parallax degeneracy affects only the inferred direction of motion. It is therefore irrelevant to any of the results discussed thus far. The direction of motion could be an interesting quantity. However, if it were to be used to measure the lens motion, one would have to make a measurement of the proper motion of the source. The latter is likely to be $\sim 10\,\mathrm{km\,s^{-1}\,kpc^{-1}} \sim 2\,\mathrm{mas\,yr^{-1}}$ in each direction and so could be roughly measured with two *HST* exposures separated by $\sim 10$ years.

Resolving the degeneracy in the direction of motion requires observing the event from a third location, not collinear with the other two (Gould 1994b). In fact, with three such observatories, one could determine the parallax from the three $t_0$ measurements alone, i.e., without any information about the impact parameters. This could be useful for the events where the lens passes well inside the source. Recall from § 2.3 that for such events $\Delta t$ is measurable but $\Delta\beta$ is not.

However, observation from three non-collinear observatories creates substantial logistical difficulties. First, in practice the third observatory would have to be either at the south pole or in the northern hemisphere. If the latter, the period each night when the bulge is observable would be short, and hence the number of northern observatories required to make routine monitoring possible would be large. Second, if three observatories are required for a measurement, the chance of weather problems is high. There would be substantial value, however, in *occasional* measurements from a third (northern) observatory. The $\beta$ and $t_0$ at this observatory are predicted by the measurements at the other two (up to a two-fold degeneracy). The measurement would therefore serve as an external check on the internal errors reported by the two southern observatories.

There is yet another form of degeneracy that could affect these measurements, uncertainty in $\omega$. Near the peak of a high-magnification event, the flux is given by

$$F(t) = \frac{F_{\max}}{[1 + (t - t_0)^2/t_{\mathrm{eff}}^2]^{1/2}}, \tag{6.1}$$

where
$$F_{\max} = \frac{F_0}{\beta}, \qquad t_{\text{eff}} = \frac{\beta}{\omega}. \tag{6.2}$$

Since $\omega$ does not appear in equation (6.1), it cannot be determined from the peak of the event. Since $\theta_e = \theta_*/\omega t_*$ and the empirically determined quantities are $\theta_*$ and $t_*$, uncertainty in $\omega$ leads to an equal uncertainty in $\theta_e$. Parallax measurements are affected similarly.

If the unlensed flux $F_0$ were known, then one could determine $\beta$ and hence $\omega$ using equation (6.2) together with the measured $F_{\max}$ and $t_{\text{eff}}$. For lensing events observed to date, one usually assumes that $F_0$ is the flux observed from the star after (or before) the event. In fact, this post-event flux may include additional light from a binary companion to the source, from the lens itself, or from a random field star.

For EMEs, the post-event flux cannot be reliably measured from the normal search observations. First, the observations are not deep enough. Second, if there are $2.5 \times 10^8$ source stars over $82\,\text{deg}^2$, then there are an average of 0.25 sources arcsec$^{-2}$, making measurements in $\sim 1''$ seeing with $0\rlap{.}''6$ pixels problematic. However, it would be straight forward to measure the post-event flux using the *HST* planetary camera. By comparing the color of the star after the event with its color at maximum one could detect or rule out the presence of additional light unless it were from a star of very similar color. Stars of similar color (whether in the bulge or the foreground) to these main-sequence sources would likely have similar or greater brightness. Such bright companions would have a significant effect on the structure of the light curve. Finally, binary companions within the Einstein ring would show up in the structure of the light curve (Griest & Hu 1992; Han & Gould 1996b). Thus, it appears likely that unlensed companions to the source could be either detected or severely constrained.

## 7. Partial Information

For transit or near-transit events with $z > z_{\max}$, it will be possible to measure $\theta_e$ but obtain only a lower limit for $\tilde{r}_e$. This limit will provide lower limits on the mass and distance through equations (1.5) and (1.6). If, for example $z_{\max} = 0.75$, then one will know that a bulge lens ($D_{\text{ls}} \lesssim 2\,\text{kpc}$) is being detected, but will have only a lower limit on its mass.

Similarly, although the fraction of nearby disk events with near transits and hence measurable proper motions is small, there will be a much larger fraction with impact parameters of several source radii that still have measurable parallaxes. In this case, there will be an upper limit on $\theta_e$ and hence on the mass and distance.

These limits, while certainly not as valuable as measurements, can be used in concert with mass measurements of other objects to constrain the overall population.

## 8. Observations Toward the LMC

The prospects for extreme microlensing toward the LMC are substantially less favorable than toward the bulge in part because there are fewer events and in part because the sources are more distant. I make a rough estimate of these prospects as follows. First, since there is less extinction toward the LMC, I assume that the observations are carried out to a limit $R \sim 23.5$ corresponding (as in the bulge) roughly to solar type stars. The actual luminosity function of the LMC at these magnitudes is unknown, so I normalize the calculation to $\sim 10^8$ source stars. Observations of the LMC can in principle be carried out all year, but during the southern winter it is observable only at the ends of the night making simultaneous follow-up by two widely separated observatories difficult or impossible. I therefore assume a 180 day summer observering season. I assume that the optical depth is $\tau \sim 2 \times 10^{-7}$ and the mean event time is $\sim 37$ days (Alcock et al. 1996b). Combining these assumptions and scaling from the previous results, I estimate there is $\sim 1$ EME toward the LMC per year. Moreover, in contrast to the bulge EMEs, there is little chance to measure proper motions for LMC EMEs because the sources are $\sim 6$ times farther away and hence 6 times smaller. It therefore appears that an EME search toward the LMC would not yield significant returns.

**Acknowledgements**: I would like to thank B. Gaudi for making several valuable suggestions. This work was supported in part by grant AST 94-20746 from the NSF.